# Electrical Switching of Tri-State Antiferromagnetic Néel Order in $\alpha$-$Fe_2O_3$ Epitaxial Films

Yang Cheng[1,*], Sisheng Yu[1,*], Menglin Zhu[2], Jinwoo Hwang[2], Fengyuan Yang[1]

[1]Department of Physics, The Ohio State University, Columbus, OH, 43210 USA

[2]Department of Materials Science and Engineering, The Ohio State University, Columbus, OH, 43212, USA

[*]These two authors contributed equally to this work.

We demonstrate non-decaying, step-like electrical switching of tri-state Néel order in Pt/$\alpha$-$Fe_2O_3$ bilayers detected by spin-Hall induced anomalous Hall effect. The as-grown Pt/$\alpha$-$Fe_2O_3$ bilayers exhibit saw-tooth switching behavior generated by current pulses. After annealing by a high pulse current, the Hall signals reveal single-pulse saturated, non-decaying, step-like switching. Together with control experiments, we show that the saw-tooth switching is due to an artifact of Pt while the actual spin-orbit torque induced antiferromagnetic switching is step-like. Our Monte-Carlo simulations explain the switching behavior of $\alpha$-$Fe_2O_3$ Néel order among three in-plane easy axes.

Spin-orbit torque (SOT) induced switching of ferromagnets (FM) by an adjacent heavy metal (HM) has raised wide interests in recent years,[1-3] where a charge current in the HM generates spins at the HM/FM interface via the spin Hall effect (SHE). Antiferromagnets (AFs) offer the advantage of no stray field, robustness against external field, THz response, and abundance of material selections.[4-12] It has been predicted that Néel SOT can be utilized to switch AF spins in picoseconds for THz operations.[13-17] Electrical switching of bi-state AF moments has been demonstrated in metallic AFs, CuMnAs and $Mn_2Au$.[18-22] For antiferromagnetic insulators (AFIs), the switching of Néel order can be achieved in HM/AFI bilayers by damping-like SOT, as shown recently in Pt/NiO bilayers with saw-tooth shaped switching, which was interpreted as that every ~1 ms current pulse can flip the AF-Néel order incrementally.[23-26] In this letter, we report the first observation of tri-state, step-like switching of Néel order in Pt(2 nm)/$\alpha$-$Fe_2O_3$(30nm) bilayers grown on $Al_2O_3$(001) substrates, which is read out by Hall resistance ($\Delta R_{xy}$) detection. Our results demonstrate that the saw-tooth $\Delta R_{xy}$ is an artifact from the Pt layer, while the SOT-induced AF switching is step-like.

Epitaxial $\alpha$-$Fe_2O_3$ films are grown on $Al_2O_3$(001) substrates using off-axis sputtering, followed by in-situ deposition of a Pt layer on $\alpha$-$Fe_2O_3$ at room temperature.[27-29] $\alpha$-$Fe_2O_3$ is a high temperature AFI with a corundum structure as shown in Fig. 1(a). The $Fe^{3+}$ moments stay in the (001) plane and stack antiferromagnetically along the $c$-axis above the Morin transition temperature (see Supplementary Materials for details).[30, 31] Figure 1(b) shows a $2\theta/\omega$ X-ray diffraction (XRD) scan of a phase-pure $\alpha$-$Fe_2O_3$(30 nm) epitaxial film on $Al_2O_3$(001), where the Laue oscillations of the $\alpha$-$Fe_2O_3$(006) peak in the inset indicate its high crystal quality. The scanning transmission electron microscopy (STEM) image of a Pt(2 nm)/$\alpha$-$Fe_2O_3$(30 nm) bilayer shown in Fig. 1(c) reveals the single-crystalline ordering of $\alpha$-$Fe_2O_3$ and the clean Pt/$Fe_2O_3$ interface.

Figure 2(a) shows the *ab*-plane of $\alpha$-$Fe_2O_3$ hexagonal lattice with three easy axes along the [210], [120], and [$1\bar{1}0$] directions.[30, 32] We pattern our Pt(2 nm)/$\alpha$-$Fe_2O_3$(30nm) bilayers into 8-leg Hall crosses, as shown in Figs. 2(b) and 2(c), where the width of the two vertical Hall terminals is 5 μm and the other six legs (60° apart) are 10 μm wide. We determine the crystallographic axes of the samples using reflection high-energy electron diffraction (RHEED) to align **E1**, **E2**, and **E3** with the [210], [120], and [$1\bar{1}0$] easy axes of $\alpha$-$Fe_2O_3$, respectively (see Supplementary Materials for details).

Hall resistances of the patterned bilayers are measured using a Physical Property Measurement System (PPMS) at 300 K unless specified otherwise. During our switching measurements, we first apply a 1-ms pulse current ($\boldsymbol{I}_p$) along one of three easy axes, wait for 30 seconds, and then measure the Hall voltage across the two vertical terminals by sending a small sensing current ($\boldsymbol{I}_s$) of 100 μA along **E2**. After a series of 10 pulses, we change the direction of $\boldsymbol{I}_p$ to another easy axis and repeat the measurement. Figure 2(d) shows $\Delta R_{xy}$ as a function of pulse count at $I_p$ = 9 mA (current density, $j = 4.5 \times 10^7$ A/cm$^2$), which exhibits clean tri-state Hall resistances at $\boldsymbol{I}_p \parallel$ **E1** (low), $\boldsymbol{I}_p \parallel$ **E2** (intermediate), and $\boldsymbol{I}_p \parallel$ **E3** (high) as $\boldsymbol{I}_p$ is switched from **E2**→**E3**→**E2**→**E1**→**E2**. This switching behavior can be understood as follows: 1) when an initial pulse current is applied along one of the three easy axes, the damping-like SOT rotates the Néel order $\boldsymbol{n}$ to align with $\boldsymbol{I}_p$,[23] 2) a small sensing current is sent along **E2** and a spin-Hall induced anomalous Hall effect (SH-AHE) voltage is measured, which reflects the orientation of $\boldsymbol{n}$, 3) after the first pulse, the subsequent 9 pulses cause essentially no change in $\boldsymbol{n}$, resulting in a plateau, 4) as $\boldsymbol{I}_p$ is changed to a new easy axis, $\boldsymbol{n}$ aligns with the new direction of $\boldsymbol{I}_p$, leading to a step-jump of $\Delta R_{xy}$. The step-like switching of the Néel order is in distinct contrast with previous reports in Pt/NiO bilayers with saw-tooth shaped $\Delta R_{xy}$.

The magnitude of the Hall resistance, $\Delta R_{xy}$(**E3**) > $\Delta R_{xy}$(**E2**) > $\Delta R_{xy}$(**E1**) arises from the relative angle of -60°, 0°, and +60° between $\boldsymbol{n}$ and $\boldsymbol{I}_s$ (which generate spins $\boldsymbol{\sigma} \perp \boldsymbol{I}_s$ in Pt vis SHE) for $\boldsymbol{I}_p$ along **E3**, **E2**, and **E1**, respectively, as expected from the angular dependence of the damping-like SOT induced SH-AHE.[23, 26, 33] To corroborate the results in Fig. 2(d), we use an independent approach to control the Néel order by an applied field ($\boldsymbol{H}$) which aligns $\boldsymbol{n} \perp \boldsymbol{H}$ via the in-plane spin-flop (SF) transition once $H$ exceeds the SF field. Figure 2(e) shows the angular dependence of $\Delta R_{xy}$ by applying an in-plane field (see Fig. 2(c) where $\alpha$ is the angle between $\boldsymbol{H}$ and **E2** or [120] crystal axis) of 0.1, 1, and 3 T, which is analogous to the planar Hall measurement in FMs. At $H \geq 1$ T, $\Delta R_{xy}$ reaches saturation and follows $\sin 2\alpha$, while at $H = 0.1$ T, it shows an irregular angular dependence, indicating that the in-plane SF transition in our $\alpha$-$Fe_2O_3$ films occurs at below 1 T with $\boldsymbol{n} \perp \boldsymbol{H}$.[33, 34] The peak-to-valley magnitude of $\Delta R_{xy}$ in Fig. 2(e) is 0.27 Ω, which gives the upper limit of Hall resistance change in Pt/$\alpha$-$Fe_2O_3$ switching measurement. The plateaus in Fig. 2(d) for E3, E2, and E1 correspond to $\alpha$ = 30°, 90°, and 150° marked in Fig. 2(e), respectively. The values of $\Delta R_{xy}$ in Fig. 2(d) are smaller as compared to the corresponding points in Fig. 2(e), and we will explain it below in Fig. 4.

Because for damping-like SOT $\propto \boldsymbol{n} \times (\boldsymbol{j} \times \hat{\boldsymbol{z}}) \times \boldsymbol{n}$, the magnitude of pulse current density $\boldsymbol{j}$ determines $\Delta R_{xy}$,[23] we measure the $I_p$ dependence of the Pt/$\alpha$-$Fe_2O_3$ samples by applying $\boldsymbol{I}_p$ along E1 and E3, as shown in Fig. 3(a). As $I_p$ increases, $\Delta R_{xy}$ changes from single-pulse saturation, step-like switching to saw-tooth shaped switching. At $I_p$ = 16 mA, there is a clear decay of $\Delta R_{xy}$ after several cycles of pulses. During the first cycle, $\Delta R_{xy}$ is ~0.3 Ω which is above the upper limit of 0.27 Ω given by Fig. 2(e). The obvious decay at $I_p$ = 16 mA has been observed in other HM/AFI switching systems, which was attributed to the decrease of switching efficiency.[19, 24]

To uncover the cause of saw-tooth switching and the decay of $\Delta R_{xy}$, we perform the same measurement using another Hall cross on the same sample in an in-plane field of 3 T applied at $\boldsymbol{H} \perp$ **E2** [Fig. 3(b)]. Since $\boldsymbol{H}$ is fixed at $\alpha = 90°$ and above the SF field, the AF moments are frozen along **E2** and no switching is expected. Surprisingly, the 3 T field has essentially no impact on $\Delta R_{xy}$ at $I_p$ = 16 mA, which remains saw-tooth like with similar magnitude. The 12 and 10 mA curves, on the other hand, shows sharp difference, becoming flat lines (no switching) in Fig. 3(b). The inset in Fig. 3(b) plots $\Delta R_{xy}$ vs. $I_p$ in a semi-log scale, exhibiting an exponential dependence. Likewise, the inset in Fig. 3(a) shows a similar plot for 0 T, where the red curve is not a fit, but the sum of exponential fit obtained in the inset of Fig. 3(b) and the linear fit from Fig. 4(c) below.

To highlight the contrast between Figs. 3(a) and 3(b), Figs. 3(c) and 3(d) show the comparison of $\Delta R_{xy}$ between the 0 and 3 T data at $I_p$ = 16 and 12 mA, respectively. In Fig. 3(c) for $I_p$ = 16 mA, there is essentially no difference between the 0 and 3 T curves despite the different AF spin configurations. In Fig. 3(d) for $I_p$ = 12 mA, the 3 T field turns the step-like $\Delta R_{xy}$ at 0 T into an essentially flat line (with a very small but non-negligible saw-tooth shape), suggesting that the step-like switching is the real AF switching while the saw-tooth feature has a different origin.

Considering the saw-tooth feature is most obvious at $I_p$ = 16 mA, we apply an even higher pulse current of 18 mA ($j = 9.0 \times 10^7$ A/cm$^2$) at zero field to anneal the 2 nm Pt layer and then redo the measurement at $I_p$ = 16 and 12 mA in a 3 T field, as shown in Figs. 3(e) and 3(f), respectively. In both cases, there is no switching and $\Delta R_{xy}$ remains flat after the annealing. We next perform the same measurement at zero field for $I_p$ = 16 and 12 mA. Figure 3(g) shows that after the annealing, the saw-tooth curve at $I_p$ = 16 mA is transformed to a step-like switching. In Fig. 3(h) for $I_p$ = 12 mA, $\Delta R_{xy}$ remains step-like while the switching becomes more square-like. This result demonstrates that the annealing dramatically changes the detected switching behavior, which we attribute to the improved stability of the Pt(2 nm) layer after the annealing.

Since the switching of Pt/$\alpha$-$Fe_2O_3$ samples becomes significantly more stable after the annealing, we can obtain a reliable $I_p$ dependence of the SOT-induced switching. Figure 4(a) shows that for the whole current range from 6 to 16 mA, $\Delta R_{xy}$ exhibits step-like switching with high stability and no detectable decay. The onset of switching occurs at $I_p$ = 6 mA or $j = 3.0 \times 10^7$ A/cm$^2$, comparable to the values for typical HM/FM systems.[2, 3] A linear-scale plot of $\Delta R_{xy}$ vs. $I_p$ shown in Fig. 4(c) exhibits a linear dependence at $I_p \geq 8$ mA. This indicates the SOT responsible for the AF switching is linearly proportional to the magnitude of $I_p$, which in turn is proportional to the SHE-generated spin accumulation at the Pt/$\alpha$-$Fe_2O_3$ interface. In addition, the fitting parameters obtained from Fig. 4(c), together with the exponential fitting to the inset in Fig. 3(b), are used to create the red curve in the inset in Fig. 3(a), which approximately agrees with the experimental data for fresh samples without the annealing.

During the switching of $\boldsymbol{n}$ from one easy axis to another, thermal fluctuation is expected[19,26] to help $\boldsymbol{n}$ overcome the potential barrier due to magnetocrystalline anisotropy. We measure the temperature ($T$) dependence of $\Delta R_{xy}$ at $I_p$ = 9 mA from 200 to 300 K at zero field as shown in Fig. 4(b), which decreases at lower temperatures as expected. Figure 4(d) shows the $\Delta R_{xy}$ vs. $T$ plot, which exhibits an exponential dependence, confirming thermally activated AF switching.[19]

Figure 4(e) shows the dependence of $\triangle R_{xy}$ on the magnitude of an in-plane field applied at $\boldsymbol{H} \perp$ **E3**, [$\alpha = 30°$, see Fig. 2(c)], which aligns $\boldsymbol{n} \parallel$ **E3** at $H$ above the SF field. As $H$ is ramped from 0 to 1 T (initial curve) and then back to 0 T, the $\triangle R_{xy}$ vs. $H$ curve exhibits a full loop in the first-quadrant, analogous to FMs. The remanence of $\triangle R_{xy}$ at $H$ = 0 T on the red curve is ~25% of the saturation value at 1 T because the $\alpha$-$Fe_2O_3$ film transitions from a single domain to multi-domains as $H$ is reduced to below the SF field. We also perform a minor loop measurement by ramping $H$ from 0 to 0.1 T and then back to 0 T, which exhibits a much smaller remanence at 0 T.

A pulse current applied along an easy axis generates SHE-induced spin accumulation near the Pt/$\alpha$-$Fe_2O_3$ interface, which acts as an effective magnetic field $\propto (\boldsymbol{j} \times \hat{\boldsymbol{z}}) \times \boldsymbol{n}$ and exerts a SOT on the Néel order to align $\boldsymbol{n}$ with $\boldsymbol{I}_p$. This is similar to an FM whose magnetization can be aligned by a magnetic field. Given the THz response of AFs[13] and that the sample temperature can be stabilized in μs,[19] a single 1-ms pulse is long enough for an AF to reach equilibrium. As a result, the percentage of Néel order switching only depends on the magnitude of $I_p$ rather than the number of pulses. Since our $\triangle R_{xy}$ is recorded using a small sensing current long after the pulse current is off, the measured signal is the remanence of $\triangle R_{xy}$, which is a fraction of the saturation value. This is analogous to the demagnetization process of FMs and can explain why $\triangle R_{xy}$ in switching measurements is much smaller than that in the field-dependence measurements shown in Fig. 2(e).

Figure 4(f) shows our Monte-Carlo simulations of the full and minor loops in Fig. 4(e) by computing the component of $\boldsymbol{n}$ along E3, $n_{E3}$, as a function of the effective magnetic field, $H_{eff}/\sqrt{2H_{k2}}$, generated by the SOT with $\boldsymbol{I}_p \parallel \mathbf{E3}$, where $H_{k2}$ is the easy-plane anisotropy field (see Supplementary Materials for details). The simulation result of SOT-induced switching qualitatively agrees with the experiment result in Fig. 4(e) induced by an external field, revealing the similarities in the control of AF spins between a magnetic field and current-induced SOT.

To uncover the reason for the saw-tooth switching, we perform the switching measurement for a Pt(2 nm) film directly deposited on $Al_2O_3$ (see Supplementary Materials), which displays the saw-tooth shaped $\triangle R_{xy}$. We speculate that the saw-tooth feature of $\triangle R_{xy}$ is due to the current-driven migration of grain boundaries in thin Pt layers. This proves that the saw-tooth feature is an artifact due to Pt and not related to the AF switching, while the actual AF switching exhibits single-pulse saturation, step-like Hall resistance, which disappears after the AF spins are "frozen" by a magnetic field. Our results point to a promising path toward controlling AF spins in insulating antiferromagnets using spin-orbit torque. Also, we proposed a criterion to separate the spin-orbit torque switching from the artifacts, where a real AF switching is unattenuated and can be suppressed by a magnetic field when exceeding the spin-flop field.

This work was primarily supported by the Department of Energy (DOE), Office of Science, Basic Energy Sciences, under Grant No. DE-SC0001304. M.L.Z. and J.H. acknowledge support (STEM) by the Center for Emergent Materials, an NSF MRSEC, under Grant No. DMR-1420451.

**Figure Captions:**

**Figure 1**. (a) Schematic of the $\alpha$-$Fe_2O_3$ hexagonal lattice with FM-aligned Fe moment in the *ab*-plane and AF coupling between adjacent *ab*-planes (oxygen atoms not shown). (b) $2\theta/\omega$ XRD scan of a 30 nm $\alpha$-$Fe_2O_3$ epitaxial film on $Al_2O_3$(001). The insert shows a zoom-in region around the $\alpha$-$Fe_2O_3$(006) peak. (c) STEM image of a Pt(2 nm)/$\alpha$-$Fe_2O_3$(30 nm) bilayer. The inset is brightness/contrast adjusted to show clear atoms in $\alpha$-$Fe_2O_3$.

**Figure 2**. (a) The *ab*-plane of $\alpha$-$Fe_2O_3$ lattice with three in-plane easy axes, [210], [120], and $[1\bar{1}0]$ labeled as E1, E2 and E3, resulting in a tri-axial anisotropy, where the double arrows represent the AF spins. (b) Optical microscopy image and (c) schematic of an eight-leg Hall cross of a Pt(2 nm)/$\alpha$-$Fe_2O_3$(30nm) bilayer, where $\alpha$ is the angle between an in-plane field and the E2 direction. (d) A sequential pulse current of $I_p$ = 9 mA is applied along one of the three easy axes (10 pulses for each segment) at 300 K and a reversible control of tri-state Hall resistance is detected by applying a 0.1 mA sensing current along E2. (e) In-plane $\alpha$ dependence of $\triangle R_{xy}$ at $H$ = 0.1, 1, and 3 T, where $\triangle R_{xy}$ saturates at $H \geq 1$ T. The gray and purple solid curves are sin2$\alpha$ fits.

**Figure 3**. Evolution of $\triangle R_{xy}$ when the pulse current is switched between E3 and E1 (10 pulses each) under (a) 0 T and (b) 3 T in-plane field applied perpendicular to E2 for a Pt(2 nm)/$\alpha$-$Fe_2O_3$(30 nm) bilayer. Insets: semi-log plots of $\triangle R_{xy}$ vs. $I_p$. The red line in inset (b) is an exponential fit, $y = (1.38 \times 10^{-11})e^{1.44x}$, and the red curve in inset (a) is given by, $y = (1.38 \times 10^{-11})e^{1.44x} + (-0.0183 + 0.00243x)$, which is the sum of the exponential fit in inset (a) here and the linear fit in Fig. 4(c). Comparison of $\triangle R_{xy}$ at 0 and 3 T with (c) $I_p$ = 16 mA and (d) $I_p$ = 12 mA for a fresh sample. Comparison of $\triangle R_{xy}$ for a fresh sample and the same sample after 18 mA annealing at (e) $I_p$ = 16 mA in a 3 T in-plane field ($\boldsymbol{H} \perp$ **E2**), (f) $I_p$ = 12 mA at 3 T, (g) $I_p$ = 16 mA at 0 T, and (h) $I_p$ = 12 mA at 0 T.

**Figure 4**. (a) Pulse current dependence of $\triangle R_{xy}$ for a Pt(2 nm)/$\alpha$-$Fe_2O_3$(30 nm) bilayer when $\boldsymbol{I}_\mathrm{p}$ is switched between E3 and E1 (10 pulses each) measured at 300 K. (b) Temperature dependence of $\triangle R_{xy}$ (between E3 and E1) at $I_\mathrm{p}$ = 9 mA. All measurements here are taken on a sample after 18 mA annealing. (c) $\triangle R_{xy}$ vs. $I_\mathrm{p}$ from (a), showing a linear dependence (red fitting line: $y = -0.0183 + 0.00243x$). (d) Semi-log plot of $\triangle R_{xy}$ vs. $T$ for $I_\mathrm{p}$ = 9 mA from (b), indicating an exponential dependence. (e) In-plane field dependence of $\triangle R_{xy}$ with $\boldsymbol{H} \perp$ **E3** [$\alpha$ = 30°, see Fig. 2(c)], which tends to align $\boldsymbol{n} \parallel$ **E3**. The field is ramped from 0 to 1 T (green), then back to 0 T (red), which corresponds to a first-quadrant full hysteresis loop. In a separate scan, $H$ is ramped from 0 to 0.1 T (green), then back to 0 T (blue), corresponding to a minor hysteresis loop. (f) Monte-Carlo simulations of the full and minor hysteresis loops of the component of $\boldsymbol{n}$ along E3 ($n_\mathrm{E3}$) as a function of the effective magnetic field due to SOT generated by a pulse current $\boldsymbol{I}_\mathrm{p} \parallel$ **E3**, which agrees with the experimental data in (e).

**References:**

1. L. Q. Liu, C. F. Pai, Y. Li, H. W. Tseng, D. C. Ralph and R. A. Buhrman, "Spin-Torque Switching with the Giant Spin Hall Effect of Tantalum," *Science* **336**, 555 (2012).
2. L. Q. Liu, O. J. Lee, T. J. Gudmundsen, D. C. Ralph and R. A. Buhrman, "Current-Induced Switching of Perpendicularly Magnetized Magnetic Layers Using Spin Torque from the Spin Hall Effect," *Phys. Rev. Lett.* **109**, 096602 (2012).
3. C. O. Avci, A. Quindeau, C. F. Pai, M. Mann, L. Caretta, A. S. Tang, M. C. Onbasli, C. A. Ross and G. S. D. Beach, "Current-induced switching in a magnetic insulator," *Nat. Mater.* **16**, 309 (2017).
4. T. Kampfrath, A. Sell, G. Klatt, A. Pashkin, S. Mahrlein, T. Dekorsy, M. Wolf, M. Fiebig, A. Leitenstorfer and R. Huber, "Coherent terahertz control of antiferromagnetic spin waves," *Nat. Photonics* **5**, 31 (2011).
5. X. Marti, I. Fina, C. Frontera, J. Liu, P. Wadley, Q. He, R. J. Paull, J. D. Clarkson, J. Kudrnovsky, I. Turek, J. Kunes, D. Yi, J. H. Chu, C. T. Nelson, L. You, E. Arenholz, S. Salahuddin, J. Fontcuberta, T. Jungwirth and R. Ramesh, "Room-temperature antiferromagnetic memory resistor," *Nat. Mater.* **13**, 367 (2014).
6. J. Železný, P. Wadley, K. Olejník, A. Hoffmann and H. Ohno, "Spin transport and spin torque in antiferromagnetic devices," *Nat. Phys.* **14**, 220 (2018).
7. V. Baltz, A. Manchon, M. Tsoi, T. Moriyama, T. Ono and Y. Tserkovnyak, "Antiferromagnetic spintronics," *Rev. Mod. Phys.* **90**, 015005 (2018).
8. W. Zhang, M. B. Jungfleisch, W. J. Jiang, J. E. Pearson, A. Hoffmann, F. Freimuth and Y. Mokrousov, "Spin Hall Effects in Metallic Antiferromagnets," *Phys. Rev. Lett.* **113**, 196602 (2014).
9. T. Jungwirth, X. Marti, P. Wadley and J. Wunderlich, "Antiferromagnetic spintronics," *Nat. Nanotechnol.* **11**, 231 (2016).
10. S. Urazhdin and N. Anthony, "Effect of polarized current on the magnetic state of an antiferromagnet," *Phys. Rev. Lett.* **99**, 046602 (2007).
11. R. Cheng, J. Xiao, Q. Niu and A. Brataas, "Spin Pumping and Spin-Transfer Torques in Antiferromagnets," *Phys. Rev. Lett.* **113**, 057601 (2014).
12. T. Satoh, R. Iida, T. Higuchi, M. Fiebig and T. Shimura, "Writing and reading of an arbitrary optical polarization state in an antiferromagnet," *Nat. Photonics* **9**, 25 (2015).
13. V. Lopez-Dominguez, H. Almasi and P. K. Amiri, "Picosecond Electric-Field-Induced Switching of Antiferromagnets," *Phys. Rev. Appl.* **11**, 024019 (2019).
14. O. Gomonay, T. Jungwirth and J. Sinova, "High Antiferromagnetic Domain Wall Velocity Induced by Neel Spin-Orbit Torques," *Phys. Rev. Lett.* **117**, 017202 (2016).
15. R. Cheng, D. Xiao and A. Brataas, "Terahertz Antiferromagnetic Spin Hall Nano-Oscillator," *Phys. Rev. Lett.* **116**, 207603 (2016).
16. R. Zarzuela and Y. Tserkovnyak, "Antiferromagnetic textures and dynamics on the surface of a heavy metal," *Phys. Rev. B* **95**, 180402 (2017).
17. H. V. Gomonay and V. M. Loktev, "Spin transfer and current-induced switching in antiferromagnets," *Phys. Rev. B* **81**, 144427 (2010).
18. P. Wadley, B. Howells, J. Zelezny, C. Andrews, V. Hills, R. P. Campion, V. Novak, K. Olejnik, F. Maccherozzi, S. S. Dhesi, S. Y. Martin, T. Wagner, J. Wunderlich, F. Freimuth, Y. Mokrousov, J. Kunes, J. S. Chauhan, M. J. Grzybowski, A. W. Rushforth, K. W. Edmonds, B. L. Gallagher and T. Jungwirth, "Electrical switching of an antiferromagnet," *Science* **351**, 587 (2016).

19. M. Meinert, D. Graulich and T. Matalla-Wagner, "Electrical Switching of Antiferromagnetic $Mn_2Au$ and the Role of Thermal Activation," *Phys. Rev. Appl.* **9**, 064040 (2018).
20. X. F. Zhou, J. Zhang, F. Li, X. Z. Chen, G. Y. Shi, Y. Z. Tan, Y. D. Gu, M. S. Saleem, H. Q. Wu, F. Pan and C. Song, "Strong Orientation-Dependent Spin-Orbit Torque in Thin Films of the Antiferromagnet $Mn_2Au$," *Phys. Rev. Appl.* **9**, 054028 (2018).
21. S. Y. Bodnar, L. Šmejkal, I. Turek, T. Jungwirth, O. Gomonay, J. Sinova, A. A. Sapozhnik, H. J. Elmers, M. Kläui and M. Jourdan, "Writing and reading antiferromagnetic $Mn_2Au$ by Néel spin-orbit torques and large anisotropic magnetoresistance," *Nat. Commun.* **9**, 348 (2018).
22. M. J. Grzybowski, P. Wadley, K. W. Edmonds, R. Beardsley, V. Hills, R. P. Campion, B. L. Gallagher, J. S. Chauhan, V. Novak, T. Jungwirth, F. Maccherozzi and S. S. Dhesi, "Imaging Current-Induced Switching of Antiferromagnetic Domains in CuMnAs," *Phys. Rev. Lett.* **118**, 057701 (2017).
23. X. Z. Chen, R. Zarzuela, J. Zhang, C. Song, X. F. Zhou, G. Y. Shi, F. Li, H. A. Zhou, W. J. Jiang, F. Pan and Y. Tserkovnyak, "Antidamping-Torque-Induced Switching in Biaxial Antiferromagnetic Insulators," *Phys. Rev. Lett.* **120**, 207204 (2018).
24. I. Gray, T. Moriyama, N. Sivadas, G. M. Stiehl, J. T. Heron, R. Need, B. J. Kirby, D. H. Low, K. C. Nowack, D. G. Schlom, D. C. Ralph, T. Ono and G. D. Fuchs, "Spin Seebeck imaging of spin-torque switching in antiferromagnetic Pt/NiO heterostructures," *arXiv:1810.03997* (2018).
25. T. Moriyama, K. Oda, T. Ohkochi, M. Kimata and T. Ono, "Spin torque control of antiferromagnetic moments in NiO," *Sci Rep* **8**, 14167 (2018).
26. L. Baldrati, O. Gomonay, A. Ross, M. Filianina, R. Lebrun, R. Ramos, C. Leveille, T. Forrest, F. Maccherozzi, E. Saitoh, J. Sinova and M. Kläui, "Mechanism of Néel order switching in antiferromagnetic thin films revealed by magnetotransport and direct imaging," *arXiv:1810.11326* (2018).
27. B. Peters, A. Alfonsov, C. G. F. Blum, S. J. Hageman, P. M. Woodward, S. Wurmehl, B. Büchner and F. Y. Yang, "Epitaxial films of Heusler compound $Co_2FeAl_{0.5}Si_{0.5}$ with high crystalline quality grown by off-axis sputtering," *Appl. Phys. Lett.* **103**, 162404 (2013).
28. A. J. Lee, J. T. Brangham, Y. Cheng, S. P. White, W. T. Ruane, B. D. Esser, D. W. McComb, P. C. Hammel and F. Y. Yang, "Metallic Ferromagnetic Films with Magnetic Damping Under $1.4 \times 10^{-3}$," *Nat. Commun.* **8**, 234 (2017).
29. F. Y. Yang and P. C. Hammel, "Topical review: FMR-Driven Spin Pumping in $Y_3Fe_5O_{12}$-Based Structures," *J. Phys. D: Appl. Phys.* **51**, 253001 (2018).
30. F. P. Chmiel, N. Waterfield Price, R. D. Johnson, A. D. Lamirand, J. Schad, G. van der Laan, D. T. Harris, J. Irwin, M. S. Rzchowski, C. B. Eom and P. G. Radaelli, "Observation of magnetic vortex pairs at room temperature in a planar $\alpha$-$Fe_2O_3$/Co heterostructure," *Nat. Mater.* **17**, 581 (2018).
31. R. Lebrun, A. Ross, S. A. Bender, A. Qaiumzadeh, L. Baldrati, J. Cramer, A. Brataas, R. A. Duine and M. Kläui, "Tunable long-distance spin transport in a crystalline antiferromagnetic iron oxide," *Nature* **561**, 222 (2018).
32. P. Chen, N. Lee, S. McGill, S. W. Cheong and J. L. Musfeldt, "Magnetic-field-induced color change in a-$Fe_2O_3$ single crystals," *Phys. Rev. B* **85**, 174413 (2012).
33. L. Baldrati, A. Ross, T. Niizeki, C. Schneider, R. Ramos, J. Cramer, O. Gomonay, M. Filianina, T. Savchenko, D. Heinze, A. Kleibert, E. Saitoh, J. Sinova and M. Klaui, "Full angular dependence of the spin Hall and ordinary magnetoresistance in epitaxial antiferromagnetic NiO(001)/Pt thin films," *Phys. Rev. B* **98**, 024422 (2018).

34. J. Fischer, O. Gomonay, R. Schlitz, K. Ganzhorn, N. Vlietstra, M. Althammer, H. Huebl, M. Opel, R. Gross, S. T. B. Goennenwein and S. Geprägs, "Spin Hall magnetoresistance in antiferromagnet/heavy-metal heterostructures," *Phys. Rev. B* **97**, 014417 (2018).

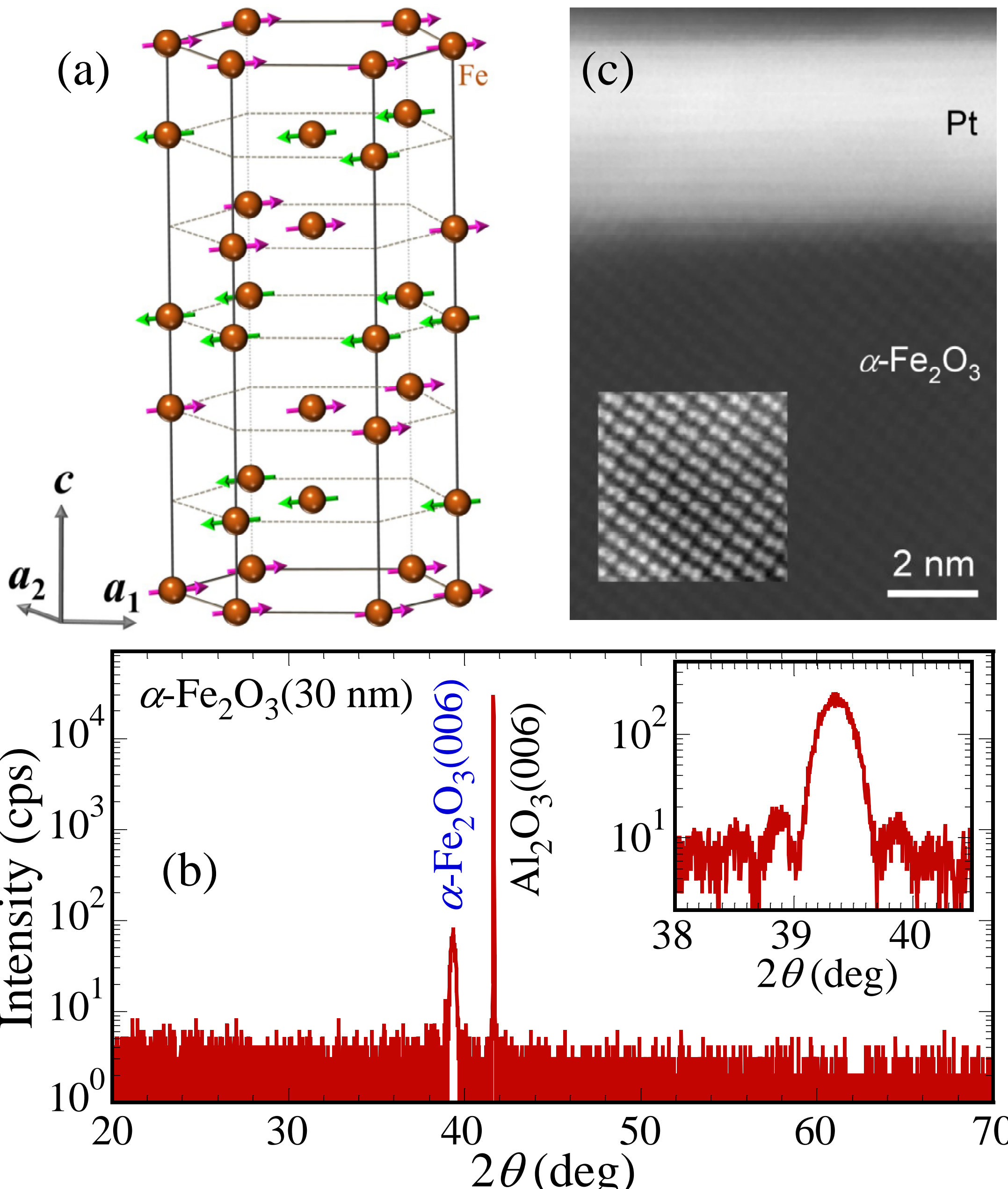

**Figure 1.**

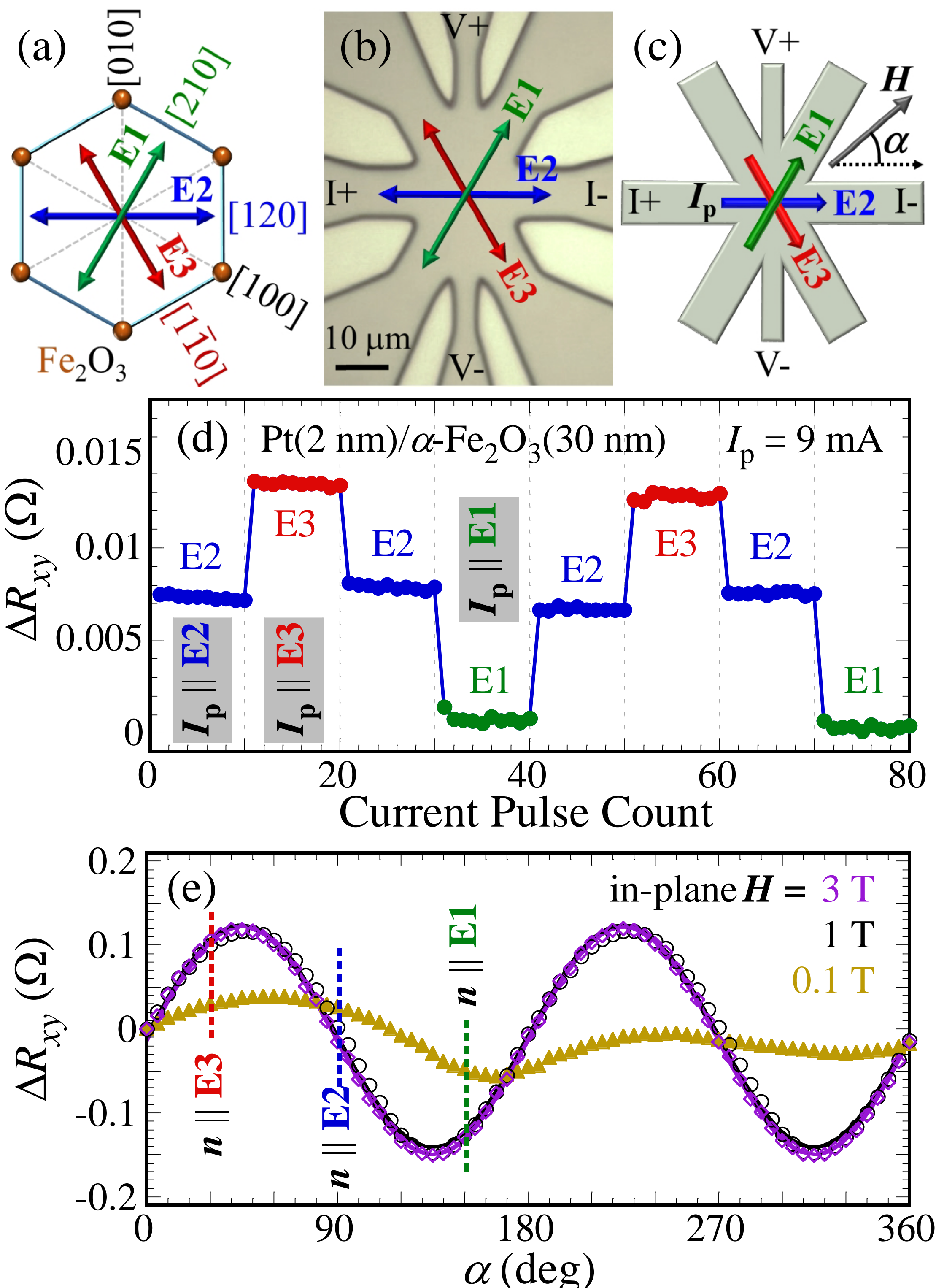

**Figure 2.**

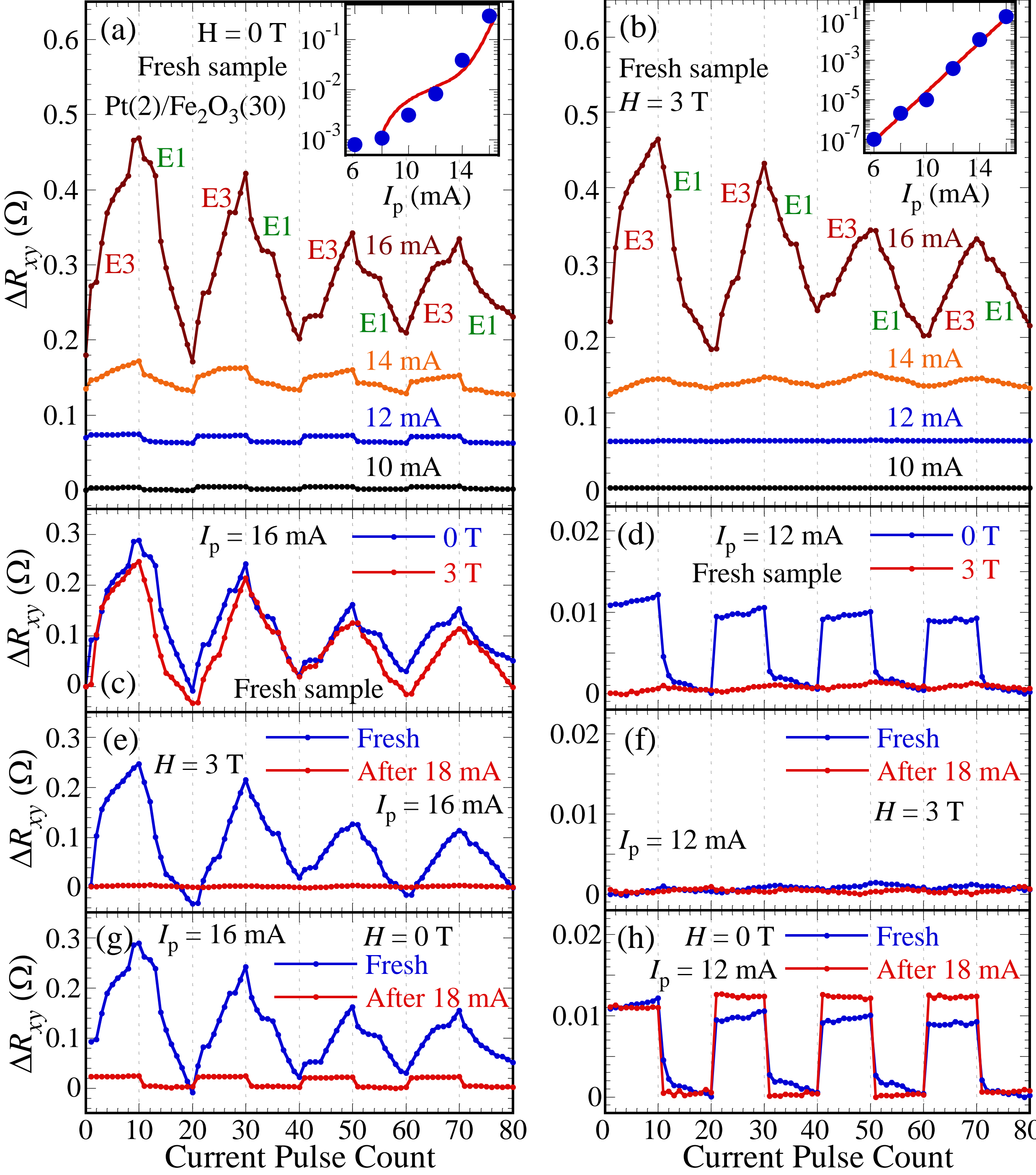

**Figure 3.**

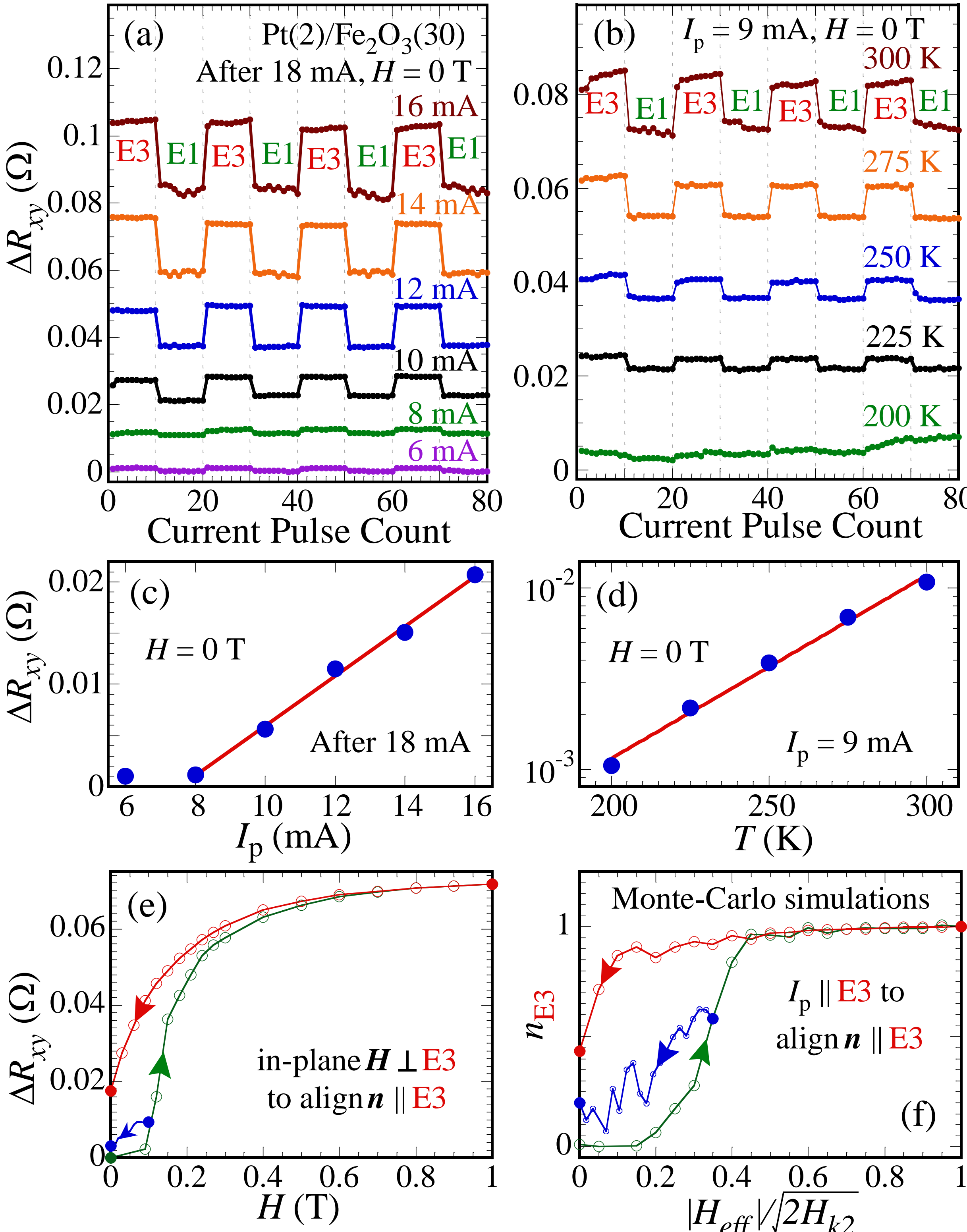

**Figure 4.**